\documentclass[11pt]{article}
\usepackage{Blois,epsfig}

\def\be{\begin{equation}}
\def\ee{\end{equation}}
\def\bea{\begin{eqnarray}}
\def\eea{\end{eqnarray}}

\def\gev{ \hbox{GeV} }

\newcommand{\lsim}{\raise.3ex\hbox{$<$\kern-.75em\lower1ex\hbox{$\sim$}}}
\newcommand{\ima}{{\mbox{Im}\,}}
\newcommand{\rea}{{\mbox{Re}\,}}
\begin{document}
\vspace*{2cm}

\vspace*{2cm}
\title{FORWARD HADRONIC SCATTERING FROM FEW GeV TO MULTI TeV WITHIN REGGE THEORY}

\author{J.R. Pel\'aez}

\address{Departamento de F\'{\i}sica Te\'orica II. Universidad Complutense. 28040 Madrid. Spain.}

\maketitle\abstracts{
The Regge description of forward hadronic scattering
is extended from $\sim1\gev$ above each threshold up to the multi TeV range.
This is achieved with a simple parametrization,
that includes mass effects,
a logarithmic growth based on an improved 
unitarity bound at intermediate energies and a separate factorization
of singularities. This parametrization can be easily implemented
for phenomenological use but also sheds light
on subdominant trajectories and the Pomeron logarithmic growth law.  
}
\section{Introduction}

In recent works\cite{Pelaez:2003eh} we have shown how it was possible to extend
down to roughly 1 GeV above threshold 
the Regge description of those combinations of total cross sections
that involved the 
Pomeron, $P'$ and $\rho$ Regge trajectories only.
In particular we provided several fits to the $(\bar{p}p+pp)$,
$(K^ +p+K^-p)$, $\pi^ \pm N$ and $\pi\pi$ cross sections.

I report here on {\it preliminary results}\cite{inprep} that
include the $a$ and $\omega$ trajectories and extend the analysis
to the total hadronic cross sections
 of $\bar{p}p$, $pp$, $\bar{p}n$, $pn$, $K^\pm p$, $K^\pm n$, $\pi^\pm N$
and $\pi\pi$, as well as the $Im F/ Re F$ ratios of  
$\bar{p}p$, $pp$, $pn$, $\pi^\pm N$ and  $K^\pm p$
forward elastic amplitudes, $F$. 
The data has been obtained from the extensive compilation of the COMPAS
group. Given the fact that the original references did not treat the
systematic uncertainties uniformly,
we have followed two fitting strategies:
First, we keep the original uncertainties as such. This allows
for an easy comparison with previous works, including
the PDG and the extensive ones of the COMPETE group\cite{Cudell:2001pn}.
However, this introduces a bias toward
those data sets, usually
the oldest, that do not provide systematic uncertainties. 
Indeed, many of these data are incompatible within their statistical
errors, and cannot be described simultaneously.
This produces an artificially large $\chi^2/d.o.f.$ 
no matter what function is fitted.
For that reason, in our second strategy, we have added a
systematic error, but only to those data
without it, of
 $0.5\%$ for $pp$,
$1\%$ for $\bar{p} p$ and $1.5\%$ for other processes.
The size of these additional errors has been chosen 
of the same order of magnitude given by 
other experiments for the same process, so that we give similar weight
to all sets without discarding any point. 
To account for different ways of combining
statistical and systematic errors, in the first strategy we have added them
in quadrature and linearly
in the second.
In addition, we use $\sigma^{total}$  data\cite{datapipi} on  $\pi^+\pi^-$, 
$\pi^-\pi^-$,$\pi^+\pi^0$, above $1.42\gev$, plus one data point per
channel reconstructed from phase shift analysis\cite{Pelaez:2003eh} 
at $1.42\gev$. If using $\pi\pi$ low energy information\cite{Pelaez:2003eh,inprep},
the $\rho$ residue and intercept come out somewhat smaller 
and larger, respectively.

The contributions to the $F_{AB\rightarrow AB}$ amplitudes from the Pomeron $P$,
the $f$ (or $P'$), $\rho$, $\omega$ and $a$ trajectories, are
\begin{eqnarray}
F_{p^\pm p}=(P_{NN}+f_{NN}+a_{NN}\mp
\omega_{NN}\mp\rho_{NN})/2, \;
F_{p^\pm n}=(P_{NN}+f_{NN}-a_{NN}\mp
\omega_{NN}\pm\rho_{NN})/2,\nonumber\\
F_{K^\pm p}=(P_{KN}+f_{KN}+a_{KN}\mp
\omega_{KN}\mp\rho_{KN})/2,\;
F_{K^\pm n}=(P_{KN}+f_{KN}-a_{KN}\mp
\omega_{KN}\pm\rho_{KN})/2,\nonumber\\
F_{\pi^\pm p}=(P_{\pi N}+f_{\pi N})/\sqrt{6}\mp\rho_{\pi N}/2,
\quad F_{\pi^\pm\pi^-}= (P_{\pi\pi}+f_{\pi \pi})/3\pm\rho_{\pi \pi}/2,
\quad
F_{\pi^0\pi^-}= (P_{\pi\pi}+f_{\pi \pi})/3, \nonumber
  \label{eq:Fsandpoles}
\end{eqnarray}
where $N=p^\pm,n$ and isospin Clebsch Gordan
coefficients have been extracted to simplify the factorization\cite{factorization}
relations $R_{AB}(\nu)=f^R_A 
f^R_B R(\nu)$, where for $R=\rho,f,a,\omega$
\begin{equation}
  \label{eq:factorization}
R(\nu)= \beta_R 
\left(\frac{1+\tau e^{-i\pi\alpha}}{\sin\pi\alpha}\right)
\nu^{\alpha_R},
\end{equation}
where $\tau$ is the signature of the trajectory.
If we want to take account of the masses, \emph{the energy dependence 
should appear through $\nu=(s-u)/2$}, instead of the usual choice of $s$.
Note that for forward scattering
 $\nu=s-m_a^2-m_b^2>s-s_{th}$. For the intercepts $\alpha_R$ we will
assume, following the QCD version of Regge theory, and the recent analysis
by the COMPETE group \cite{Cudell:2001pn}, 
that the $f/a$ and $\rho/\omega$ trajectories
are degenerate, that is, that $\alpha_a=\alpha_f$ and 
$\alpha_\omega=\alpha_\rho$.
In addition, some factors are redundant and can be absorbed 
by setting $f^R_\pi=1$,
for $R=P,f,\rho$ and $\beta_R=1$ for $R=a,\omega$.
This choice eases the notation and the 
comparison with previous works \cite{Pelaez:2003eh}.

For the Pomeron, we propose the use of a ``constant plus logarithm'', i.e.,
\begin{equation}
  \label{eq:pomeron}
P_{AB}=C_{AB}+L_{AB}, \quad\ima P(\nu)=\nu \,\left(\beta_P+A\log^2
\left[
\frac{\nu-\nu_{th}}{\nu_1\log^{7/2}(\nu/\nu_2)}
\right]
\right),
\end{equation}
where the logarithmic law follows an improved
unitarity bound\cite{Yndurain:1973rx}.
The choice $\nu-\nu_{th}$ yields slightly better fits. Note that
$\nu_{th}$ is $\nu$ at the branch point of the right cut for each amplitude.
Indeed, it has been confirmed \cite{Cudell:2001pn} that the 
$\sigma^{tot}$ growth and the
$\rea F/\ima F$ ratios, are better described with
$s\log s$ or, slightly better \cite{Cudell:2001pn}, 
with $s\log^2 s$ terms
rather than with $\alpha_P>1$.
Our Eq.(\ref{eq:pomeron})
grows faster than $s\log s$
but slower at intermediate energies than the $s\log^2 s$ 
Froissart bound, which is nevertheless 
recovered at very high $s$. 
The recent generalization of the ``factorization theorem''\cite{Cudell:2002ej} 
requires each
singularity to factorize separately. 
Thus, as a first approximation, and for the Pomeron, we use separated factors: $f_A^C$ and $f_A^L$.

Finally, the real parts of amplitudes 
are obtained from the dispersive representation, whereas total cross sections
are given by:
\begin{eqnarray}
\sigma_{ab}=4\pi^ 2 \hbox{Im}\, F_{a+b\rightarrow a+b}(s,0)
/\lambda^{1/2}(s,m_a^2,m_b^2), \quad
\lambda(s,m_a^2,m_b^2)=s^2+(m_a^2-m_b^2)^2-2s(m_a^2+m_b^2)
\nonumber
\label{eq:totcrossamp}
\end{eqnarray}
However, $\lambda$
is usually approximated by $s^2$. Only very recently\cite{Cudell:2003ci}
a slight improvement in $\chi^2/dof$ has been found
by keeping the whole $\lambda$, instead of $s^ 2$, down to $\sqrt{s}=5\,$GeV. 
Let us remark that, if $E_{kin}\simeq 1\,$GeV,
the effect of using $s^2$, instead of $\lambda$,
yields a 30\% overestimation for $NN$.

\section{Results}

We show in Table 1 the parameters of the fits to data
for different strategies and different $E_{kin}^{min}$, including
the nominal uncertainty obtained from the $\chi^2/dof$ minimization routine
MINUIT. Since the parameters are very strongly correlated 
these errors should be 
taken only nominally around the central values provided for strategy 2. 
Since there are flat directions in parameter space it is possible to get
somewhat different parameter sets with almost the same $\chi^2/dof$ but
beyond those nominal errors. An indication of the systematic 
error {\it in the determination
of single parameters} can be obtained from the difference between
strategies.
\vspace*{.2cm}

\hspace*{-.5cm}
\begin{tabular}{|c|c|c||c|}
\hline
& strategy 2& strategy 1&Minuit\\
$E_{kin}^{min}$& 1-1.3 GeV& 1-1.3 GeV&errors\\ \hline
$\beta_P$ &0.746&0.937&0.003\\
$f_N^P$ &1.792&1.705&0.007\\
$f_K^P$ &0.731&0.714&0.004\\
$A$ &0.043&0.050&0.001\\
$\nu_1$ &0.0005&0.001&0.0001\\
$\nu_2$ &0.676&0.633&0.001\\
$f_N^{log}$ &1.02&0.993&0.001\\
$f_K^{log}$ &0.723&0.733&0.012\\ 
$\beta_f$ &1.70&1.77&0.014\\
$f_N^f$ &1.78&1.75&0.01\\ \hline
\multicolumn{3}{c}{ }
\end{tabular}
\begin{tabular}{|c|c|c||c|}
\hline
& strategy 2& strategy 1&Minuit \\
$E_{kin}^{min}$& 1-1.3 GeV& 1-1.3 GeV&errors\\ \hline
$f_K^f$ &0.30&0.32&0.01\\
$\alpha_f$ &0.646&0.640&0.002\\
$f_N^a$ &-0.24&0.25&0.04\\
$f_K^a$ &-0.55&0.5&0.1\\ 
$\beta_\rho$ &1.28&1.34&0.11\\
$f_N^\rho$ &0.51&0.46&0.04\\
$f_K^\rho$ &0.49&0.54&0.04\\
$\alpha_\rho$ &0.464&0.464&0.003\\
$f_N^\omega$ &1.97&1.98&0.015\\
$f_K^\omega$ &0.66&0.65&0.01\\
\hline
$\sigma_{LHC}$&109 mb&110 mb&1mb\\
\hline
\end{tabular}

{\footnotesize {\bf Table 1.} Fit parameters with different strategies.
The Minuit errors are {\it just statistical}, and nominal, since the parameters are strongly correlated
and can only be used with the central values of {\it strategy 2}. 
} 
\label{elesln}

\begin{minipage}{\textwidth}
\vspace*{.2cm}
\begin{center}
\begin{tabular}{|c|c|c|c|c|}
\hline
$E_{kin}^{min}$ (GeV)&1-1.3& 1.5 & 2 & 3 \\ \hline
$\#$ data points& 1186 & 1002 &  895 & 768 \\ \hline
\hline Parametrization 
&
\multicolumn{4}{|c|}{$\chi^2$/d.o.f. for strategy 2 / 1}  \\ \hline
Ours &0.85/1.56&0.63/1.14&0.57/1.05&0.52/0.95\\ \hline
$\nu_1\equiv0.01$ GeV$^2$ &0.85/1.57&0.63/1.26&0.58/1.06&0.52/0.97\\
\hline\hline
powers of $s^\alpha$ &1.58/2.87&1.16/2.11&0.99/1.80& 0.78/1.42\\\hline \hline 
\multicolumn{5}{|c|}{Pomeron logarithmic term}  \\
\hline 
$\nu\, log (\nu)$ &1.01/1.83&0.69/1.26&0.59/1.09 &0.52/0.97\\
\hline
$\nu\, log (\nu-\nu_{th})$ &1.03/1.83&0.69/1.27&0.59/1.12 &0.52/0.98\\
\hline $\nu\, log^2 (\nu)$ &0.97/1.79&0.68/1.24&0.59/1.10 &0.52/0.95\\
\hline
$\nu\, log^2 (\nu-\nu_{th})$ &0.91/1.68&0.65/1.18&0.58/1.06 &0.52/0.95\\
\hline \hline 
\multicolumn{5}{|c|}{Factorization of Pomeron logarithms}  \\
\hline 
$f_a^L\equiv1$ (as PDG) &0.92/1.70&0.66/1.23&0.59/1.10 &0.54/1.01\\
\hline 
$f_a^C=f_a^L$ &0.89/1.67&0.64/1.41&0.60/1.14&0.58/1.02\\
\hline
\end{tabular}

\vspace{.2cm}
{\footnotesize {\bf Table 2.} $\chi^2$/d.o.f. for several
  $E_{kin}^{min}$ and different modifications of our parametrization.
}
\vspace*{.2cm}
\end{center}
\end{minipage}

In Table 2 we compare the $\chi^2/d.o.f.$ of parameterizations where
we have not implemented some of our suggestions. Let us recall that these
are: i)  
 using $\nu$ instead of $s$, ii) the Eq.(\ref{eq:pomeron}),
iii) the separate factorization, i) and the correct flux factor.
For each one of them
 {\it we find an improvement in $\chi^2/d.o.f.$}, with both fitting strategies.

In Figure 1 we show the curves obtained from our parametrization
including, as gray bands,
 the nominal uncertainties from strategy 2. The dashed lines correspond to the PDG2004 parametrization \cite{PDG}, 
valid above $\sqrt{s}=5$ GeV,
but naively extrapolated below.
 We see that the simple parametrization reported here 
describes remarkably well 20 observables
extending from several TeV down to $\sim 1\gev$ above the threshold
of each reaction. Further details will be given in 
a forthcoming publication\cite{inprep}, but apart from establishing the logarithmic
growth of the Pomeron, we hope it could be easily used for 
dispersive studies in hadronic physics that involve integrals
from the resonance region to infinity.

\hspace*{-.7cm}
\begin{minipage}{\textwidth}
\vspace{.3cm}
\begin{center}
  \psfig{figure=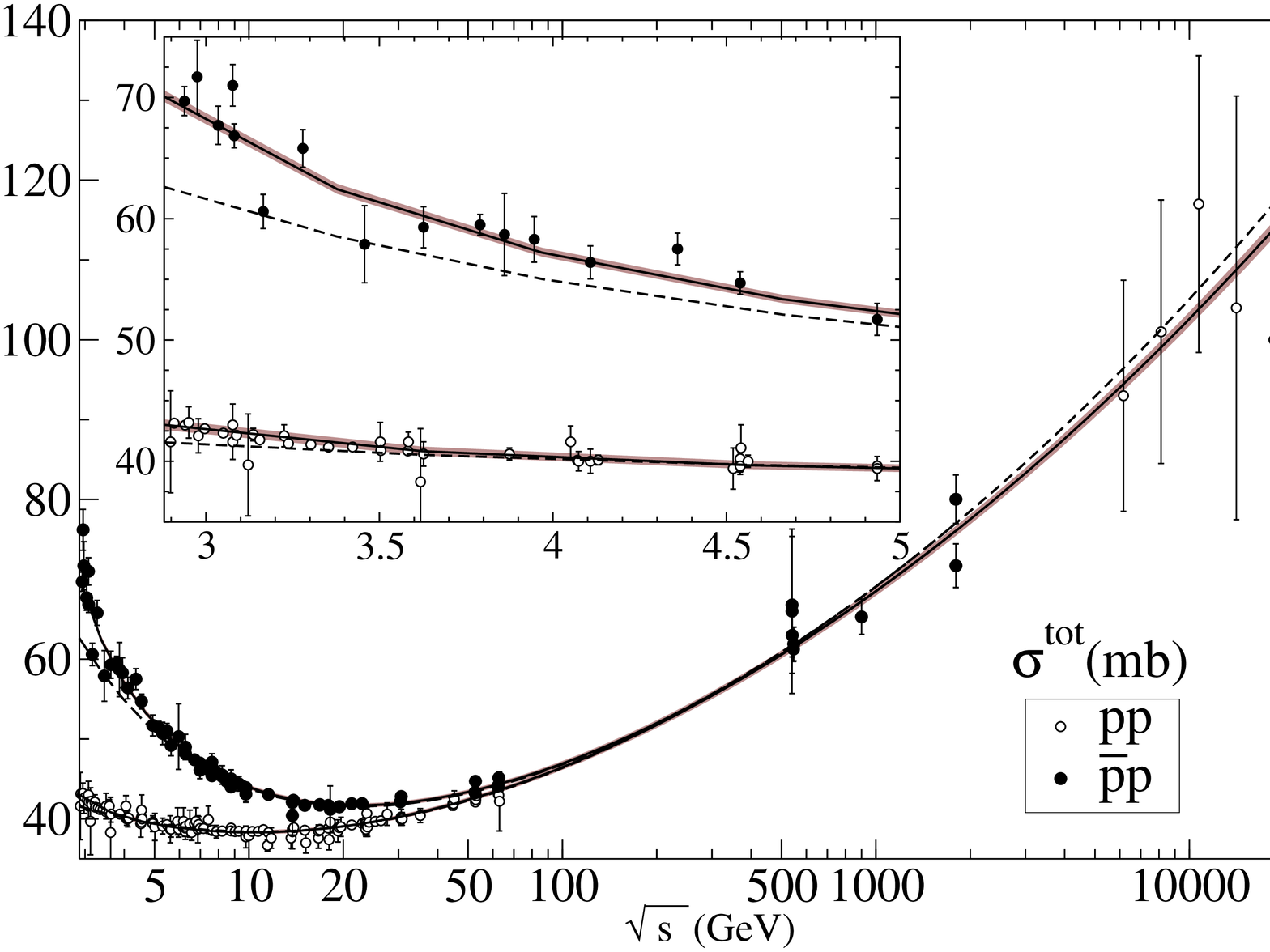,height=6.45cm,angle=-90}
  \psfig{figure=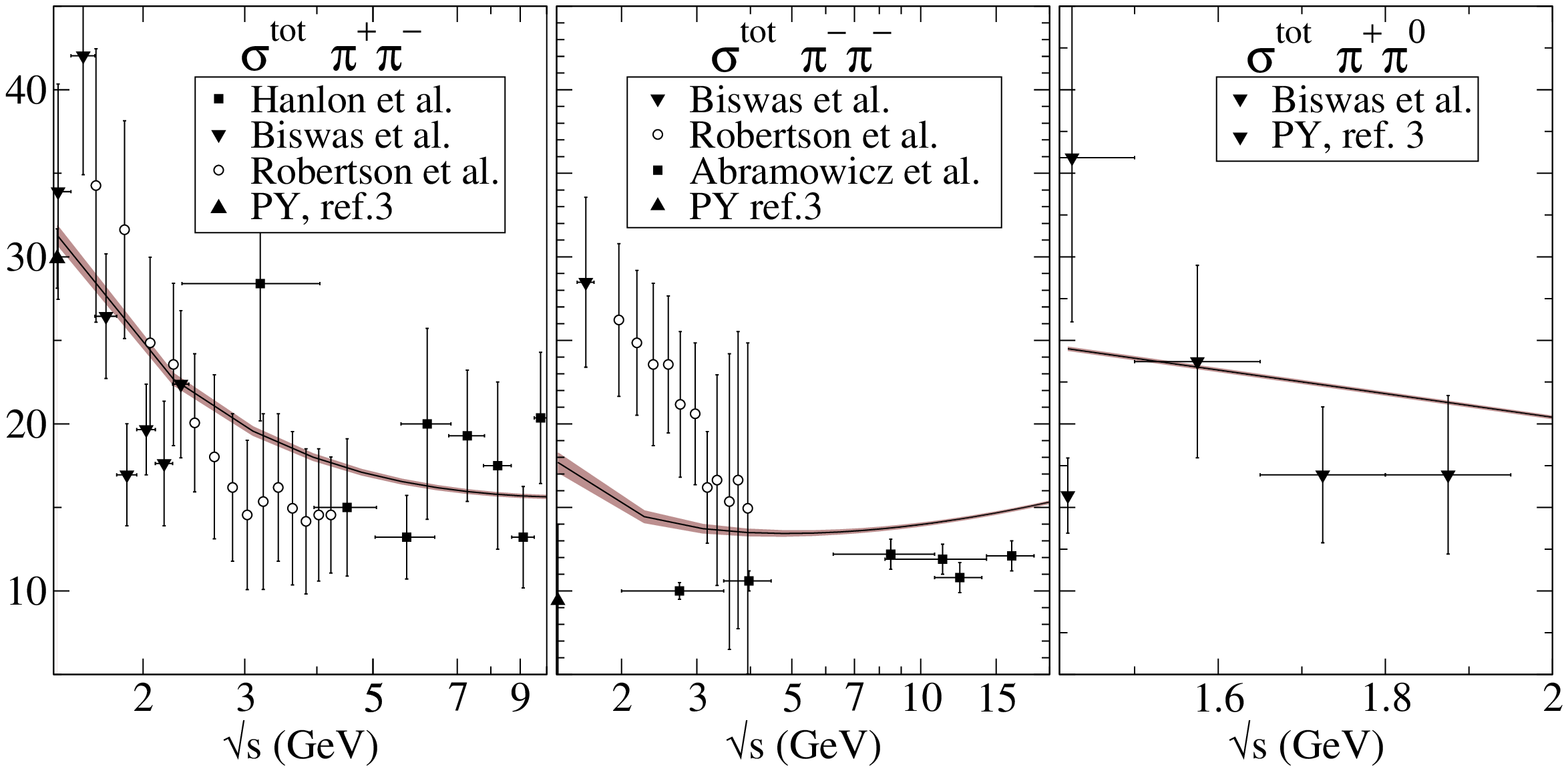,height=9.4cm,angle=-90}
  \psfig{figure=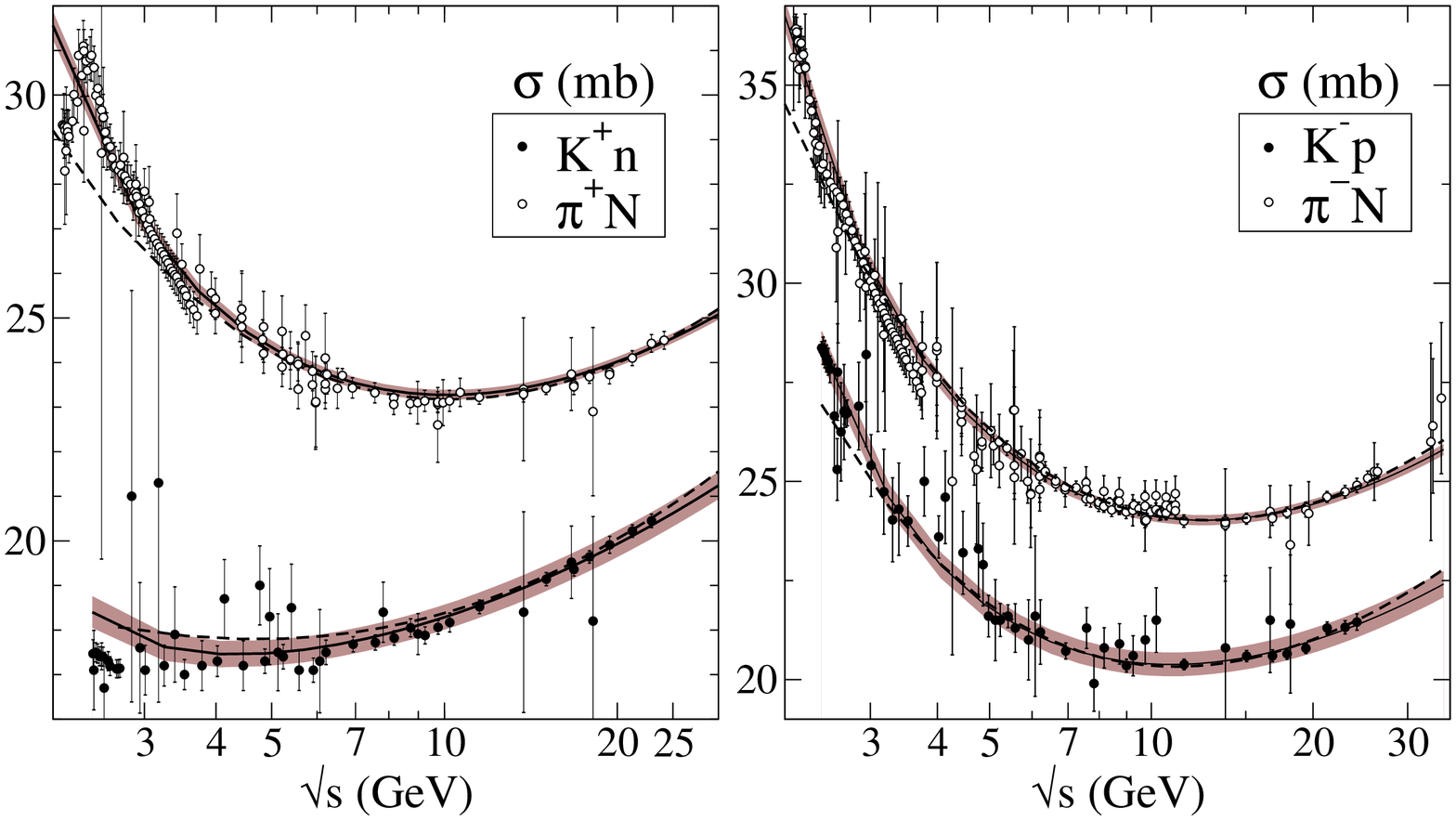,height=7.9cm,angle=-90}
  \psfig{figure=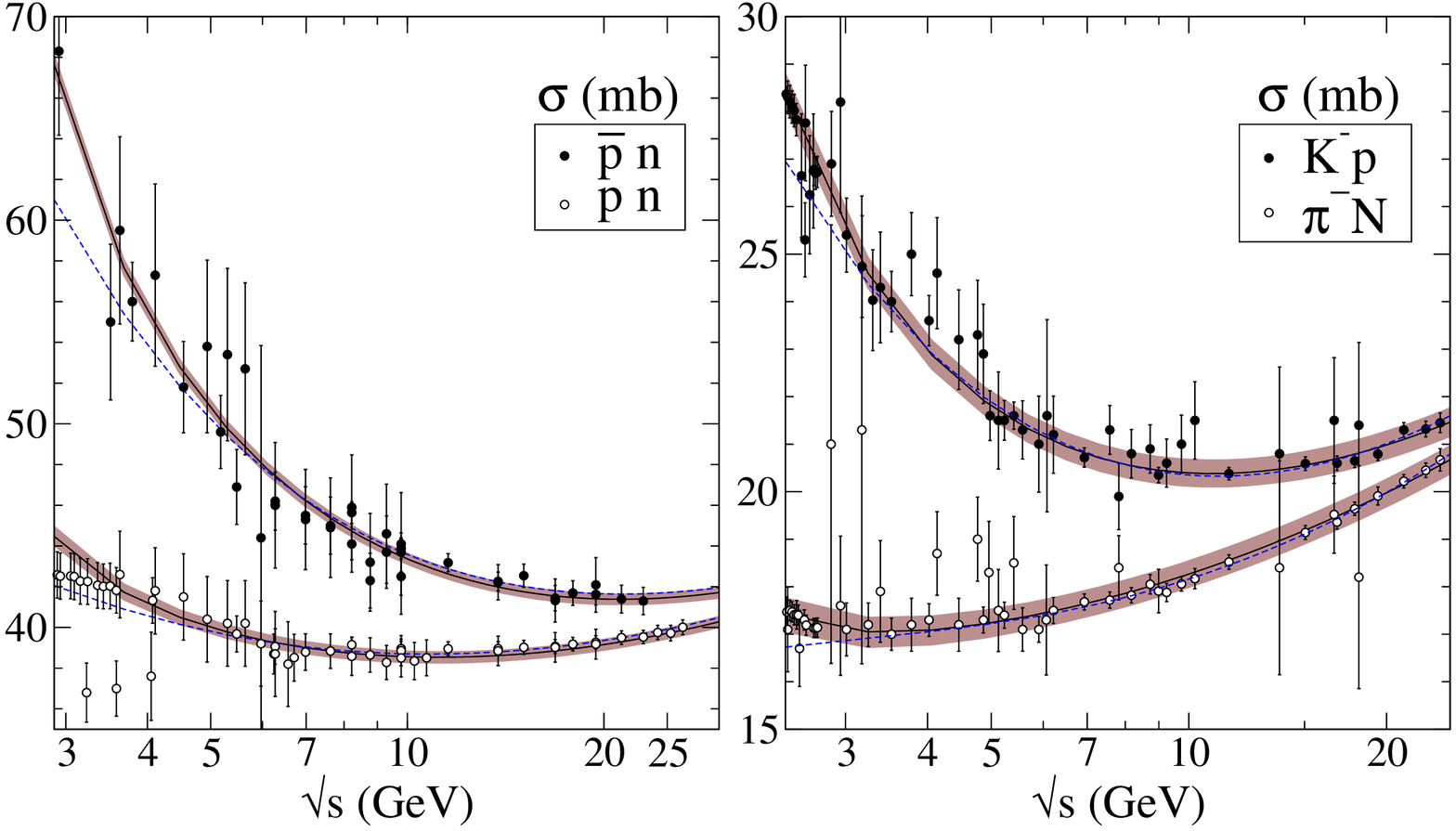,height=7.9cm,angle=-90}
  \psfig{figure=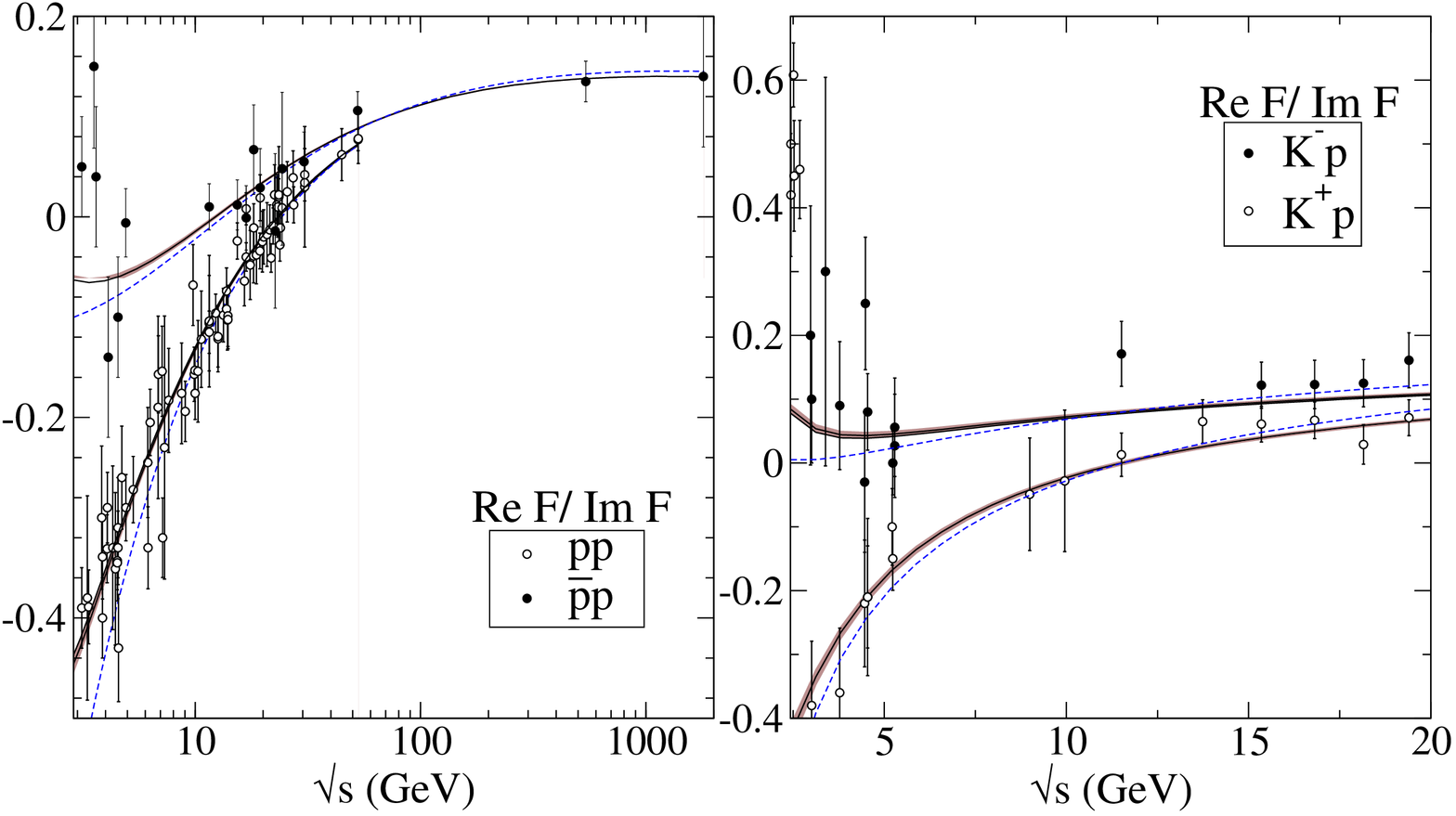,height=7.9cm,angle=-90}
  \psfig{figure=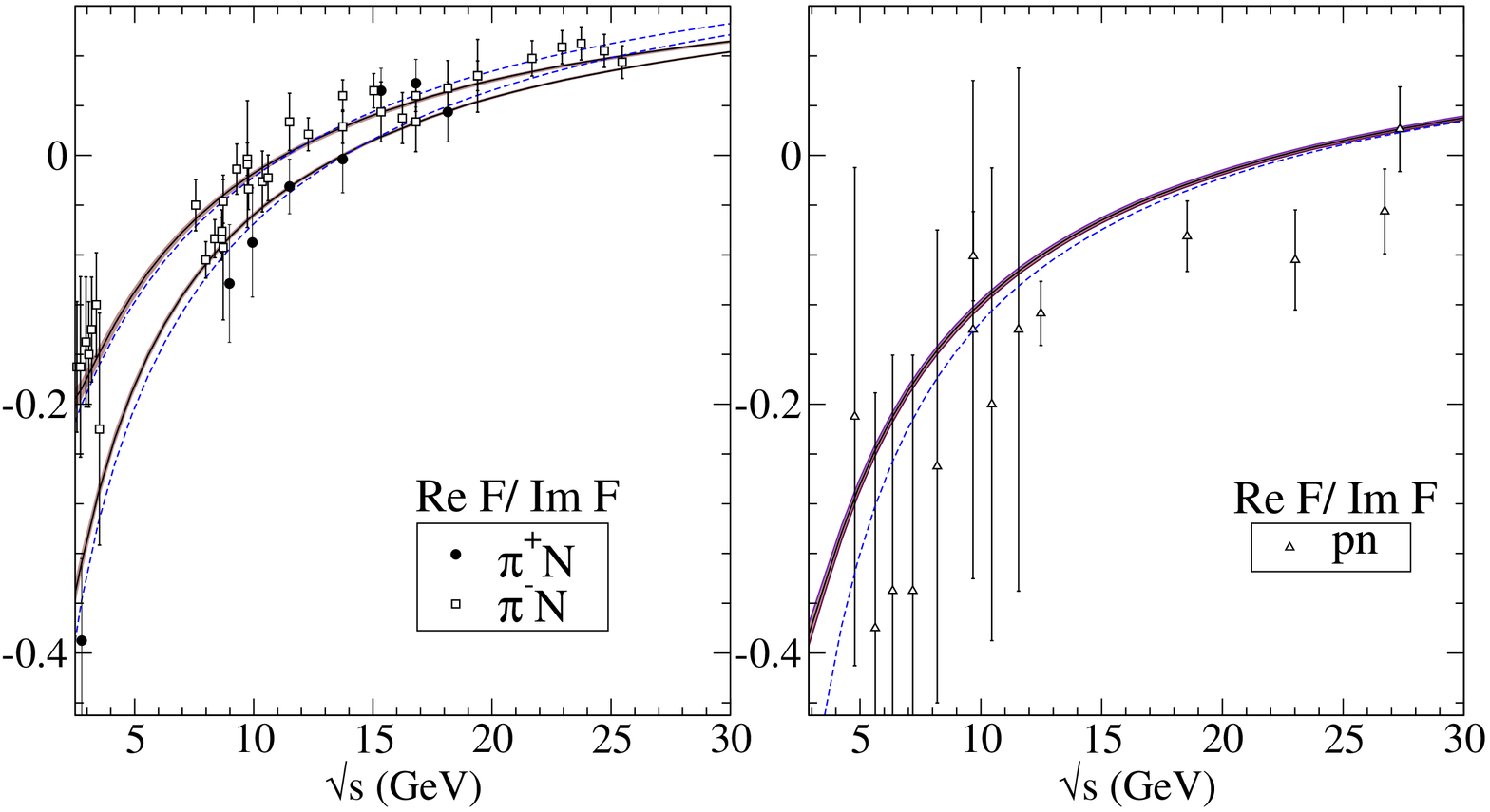,height=7.9cm,angle=-90}
\end{center} {\footnotesize Figure 1: Results from our fit down
to 1 GeV above threshold.
Total $NN$, $\pi\pi$, $\pi^ \pm N$,$K^\pm p$ and $K^\pm n$ 
cross sections as a function
of $\sqrt{s}$, with detail of $pp$ and $\bar pp$ at low energies.
In the bottom row  we show the results for $Re F/Im F$. 
The bands cover the nominal uncertainties
in the parameters. }
\vspace{.3cm}
\end{minipage}

\end{document}